\documentclass[12pt,a4paper]{article}
\usepackage{epsfig}
\usepackage{rotating}
\begin{document}
\pagestyle{empty}
\begin{titlepage}
\vspace{1. truecm}
\begin{center}
\begin{Large}
{\bf Polarization to Probe an Extra Neutral Gauge Boson at $e^+e^-$ Linear
Collider\footnote
{Talk given at the International School-Seminar "The Actual 
Problems of Particle Physics", Gomel,  July 30 -- August 8, 1999
}
}
\end{Large}

\vspace{2.0cm}

{\large  A. A. Pankov}
\\[0.3cm]

Department of Physics, Technical University, Gomel, Belarus

\end{center}
\vspace{1.0cm}

\begin{abstract}
\noindent
The sensitivity to the $Z^\prime$ couplings of the processes 
$e^+e^-\to l^+l^-,\bar{b}b$ and $\bar{c}c$ at the linear collider with 
$\sqrt{s}=500\, {\rm GeV}$ with initial beam polarization, for typical 
extended model examples are studied. 
To this aim, the suitable integrated, polarized, observables 
directly related to the helicity cross sections that carry information on the 
individual $Z^\prime$ chiral couplings to fermions are used. We discuss the  
derivation of separate, model-independent limits on the couplings in the case 
of no observed indirect $Z^\prime$ signal within the expected experimental 
accuracy. In the hypothesis that such signals were, 
indeed, observed we assess the expected accuracy on the numerical 
determination of such couplings and the consequent range of $Z^\prime$ masses 
where the individual models can be distinguished from each other as the source 
of the effect.  

\noindent

\vspace*{3.0mm}

\noindent
\end{abstract}
\end{titlepage}

\pagestyle{plain}

\section{Introduction}
Extra neutral gauge bosons are a feature of many models of physics beond the
Standard Model (SM). If discovered they would represent irrefutable proof of
new physics, most likely that the SM gauge group must be extended 
\cite{Langacker, Altarelli0}. The search for   
the $Z^\prime$ is included in the physics programme of all the present and 
future high energy collider facilities. In particular, the strategies for the 
experimental determination of the $Z^\prime$ couplings to the ordinary SM 
degrees of freedom, and the
relevant discovery limits, have been discussed in the large, and still growing, 
literature on this subject \cite{Langacker}-\cite{Leike}. \par 
Taking into account the limit $M_{Z^\prime}> 600-700\, {\rm GeV}$ from 
`direct' searches at the Tevatron \cite{Tevatron}, only `indirect' (or 
virtual) manifestations of the $Z^\prime$ can be expected at LEP2 
\cite{lep2} and at the planned $e^+e^-$ linear collider (LC) with CM energy 
$\sqrt s=500$ GeV \cite{nlc, Accomando}. 
\par 
Such effects would be represented by deviations from the calculated SM
predictions of the measured observables relevant to the different processes. 
In this regard, of particular interest for the LC is the annihilation into 
fermion pairs
\begin{equation} e^++e^-\to \bar{f}+f\hskip 2pt, 
\label{proc}
\end{equation}
that gives information on the $Z^\prime ff$ interaction. 
\par 
In the case of no observed signal within the experimental accuracy, limits on
the $Z^\prime$ parameters to a conventionally defined confidence level can be 
derived, either from a general analysis taking into account the full set of 
possible $Z^\prime$ couplings to fermions, or in the framework of specific 
models where characteristic relations among the couplings strongly reduce the 
number of independent free parameters. Clearly, completely model-independent 
limits can 
result only in the optimal situation where the different couplings can be
disentangled, by means of suitable observables, and analysed independently so 
as to avoid potential cancellations. The essential role of the initial 
electron beam polarization has been repeatedly emphasized in this regard, and 
the potential of the linear collider along these lines has been extensively 
reviewed, e.g., in Refs.~\cite{Godfrey, Leike}.
\par 
The same need of a procedure to disentangle the different $Z^\prime$ couplings 
arises in the case where deviations from the SM were experimentally observed. 
Indeed, in this situation, the numerical values of the individual couplings 
must be extracted from the measured deviations in order to identify the source
of these effects and to make tests of the various theoretical models. 
\par 
In what follows, we discuss the role of two particular, polarized, variables  
$\sigma_+$ and $\sigma_-$ in the analysis of the $Z^\prime ff$ interaction 
from both points of view, namely, 
the derivation of model-independent limits in the case of no observed deviation 
and the 
sensitivity to individual couplings and model identification in the hypothesis
of observed deviations.
\par 
These observables could directly distinguish the helicity cross sections of
process (\ref{proc}) and, therefore, depend on a minimal number of independent 
free parameters (basically, the product of the $Z^\prime$ chiral couplings to 
electrons and to the fermionic final state). They have been previously 
introduced to study $Z^\prime$ effects at LEP2 (no polarization 
there) \cite{Pankov1,Babich} and manifestations of four-fermion contact
interactions at the LC \cite{Pankov2}. Here, we extend the analysis of 
\cite{Pankov1,Babich,Pankov3} to the case of the LC with polarized beams. For 
illustration, we will explicitly consider a specific class of 
$E_6$-motivated models and of Left-Right symmetric models. 
 
\section{Polarized observables}
The polarized differential cross section for process (\ref{proc}) with 
$f\neq e,\hskip 2pt t$ is given in Born approximation by the $s$-channel 
$\gamma$, $Z$ and $Z^\prime$ exchanges. Neglecting $m_f$ with respect
to the CM energy $\sqrt s$, it has the form 
\begin{equation}
\frac{d\sigma}{d\cos\theta}
=\frac{3}{8}
\left[(1+\cos\theta)^2\hskip 2pt {\tilde\sigma}_+ 
+(1-\cos\theta)^2\hskip 2pt {\tilde\sigma}_-\right], 
\label{cross}
\end{equation}
where, in terms of helicity cross sections
\begin{equation}
{\tilde\sigma}_{+}=\frac{1}{4}\,
\left[(1+P_e)(1- P_{\bar{e}})\,\sigma_{RR}
+(1-P_e)(1+P_{\bar{e}})\,\sigma_{LL}\right], \label{s+}\end{equation}
\begin{equation}
{\tilde\sigma}_{-}=\frac{1}{4}\,
\left[(1+P_e)(1- P_{\bar{e}})\,\sigma_{RL}
+(1-P_e)(1+ P_{\bar{e}})\,\sigma_{LR}\right], \label{s-} \end{equation}
with ($\alpha,\beta=L,R$)
\begin{equation} 
\sigma_{\alpha\beta}=N_C\hskip 2pt\sigma_{pt}\hskip
2pt\vert A_{\alpha\beta}\vert^2.\label{helcross}\end{equation}
In these equations, $\theta$ is the angle between the initial electron and the 
outgoing fermion in the CM frame; $N_C$ the QCD factor $N_C\approx 
3(1+\frac{\alpha_s}{\pi})$ for quarks and  $N_C=1$ for
leptons, respectively; $P_e$ and $P_{\bar e}$ are the degrees 
of longitudinal electron and positron polarization;  
${\sigma_{\rm pt}\equiv\sigma(e^+e^-\to\gamma^\ast\to l^+l^-)}={
(4\pi\alpha_{e.m.}^2)/(3s)}$; $A_{\alpha\beta}$ are the helicity amplitudes.
\par
According to Eqs.~(\ref{s+}) and (\ref{s-}), the cross sections for 
the different combinations of helicities, that carry the information on the 
individual $Z^\prime ff$ couplings, can be disentangled {\it via} 
the measurement of  
${\tilde\sigma}_{+}$ and ${\tilde\sigma}_{-}$ with different choices of the 
initial beams polarization. Instead, the total cross section and the 
forward-backward asymmetry, defined as: 
\begin{equation}
\label{conventional}
\sigma=\sigma^{\rm F}+\sigma^{\rm B};\qquad   
A_{\rm FB}=(\sigma^{\rm F}-\sigma^{\rm B})/
\sigma,\end{equation}
with 
$\sigma^{\rm F}
=\int_{0}^{1}(d\sigma/d\cos\theta)d\cos\theta$ and 
${\sigma^{\rm B}
=\int_{-1}^{0}(d\sigma/d\cos\theta)d\cos\theta}$, depend on linear 
combinations of all helicity cross sections even for longitudinally polarized 
initial beams. One can notice the relation 
\begin{equation}
\label{relation}
{\tilde\sigma}_{\pm}=0.5\,\sigma
\left(1\pm\frac{4}{3}A_{\rm FB}\right)=\frac{7}{6}\sigma_{\rm F,B}-
\frac{1}{6}\sigma_{\rm B,F}. \end{equation}
\par 
Alternatively, one can directly project out ${\tilde\sigma}_+$ and 
${\tilde\sigma}_-$ from Eq.~(\ref{cross}), as differences of integrated 
observables. To this aim, we define $z^\ast>0$ such that 
\begin{equation}
\left(\int_{-z^*}^1-\int_{-1}^{-z^*}\right)\left(1-\cos\theta\right)^2
d\cos\theta=0. \end{equation}
Numerically, $z^*=2^{2/3}-1=0.59$, corresponding to 
$\theta^*=54^\circ$,\footnote{In the case of a reduced angular range 
$\vert\cos\theta\vert<c$, one has ${z^*=(1+3c^2)^{1/3}-1}$.} and for this 
value of $z^\ast$:
\begin{equation} 
\left(\int_{-z^*}^1-\int_{-1}^{-z^*}\right)\left(1+\cos\theta\right)^2
d\cos\theta=8\hskip 1pt\left(2^{2/3}-2^{1/3}\right). \end{equation}
From Eq.~(\ref{cross}) one can easily see that the observables   
\begin{eqnarray}
\label{sigma+}
\sigma_+&\equiv&\sigma_{1+}-\sigma_{2+}=
\left(\int_{-z^*}^1-\int_{-1}^{-z^*}\right)
\frac{d\sigma}{d\cos\theta}\ d\cos\theta, \\
\label{sigma-}
\sigma_-&\equiv&\sigma_{1-}-\sigma_{2-}=
\left(\int_{-1}^{z^*}-\int_{z^*}^1\right)
\frac{d\sigma}{d\cos\theta}\ d\cos\theta
\end{eqnarray}
are such that 
\begin{equation}
{\tilde\sigma}_{\pm}=\frac{1}{3\hskip 1pt\left(2^{2/3}-2^{1/3}\right)}
\hskip 1pt 
\sigma_{\pm}=1.02\hskip 1pt{\sigma}_{\pm}.
\label{tildesigma}\end{equation}
Therefore, for practical purposes one can identify 
$\sigma_{\pm}\cong{\tilde\sigma}_{\pm}$ to a very good 
approximation. Although the two definitions are practically equivalent from 
the mathematical point of view, in the next Section we prefer to use 
$\sigma_{\pm}$, that are found more convenient to discuss the expected 
uncertainties and the corresponding sensitivities to the $Z^\prime$ couplings. 
Also, it turns out numerically that $z^\ast=0.59$ in (\ref{sigma+}) and 
(\ref{sigma-}) maximizes the statistical significance of the results.
\par 
The helicity amplitudes $A_{\alpha\beta}$ in Eq.~(\ref{helcross}) can be 
written as
\begin{equation}
A_{\alpha\beta}=(Q_e)_\alpha(Q_f)_\beta+g_\alpha^e\,g_\beta^f\,\chi_Z+
{g^{\prime}}^e_{\alpha}\,{g^{\prime}}^f_{\beta}\,\chi_{Z^\prime},
\label{amplit}
\end{equation}
in the notation where the general neutral-current interaction is written 
as
\begin{equation} 
-L_{NC}=eJ_{\gamma}^{\mu} A_{\mu}+g_ZJ_{Z}^{\mu} Z_{\mu}
+g_{Z^\prime}^{\mu}J_{Z^\prime}^{\mu} Z_{\mu}^{\prime}. \label{notation} 
\end{equation}
Here, $e=\sqrt{4\pi\alpha_{e.m.}}$; $g_Z=e/s_W c_W$ ($s_W^2=1-c_W^2\equiv
\sin^2\theta_W$) and $g_{Z^\prime}$ are the $Z$ and $Z^\prime$
gauge couplings, respectively. Moreover, in (\ref{amplit}),   
$\chi_i=s/(s-M^2_i+iM_i\Gamma_i)$ are the gauge boson propagators with $i=Z$ 
and $Z'$, and the $g$'s are the left- and right-handed fermion
couplings. The fermion currents that couple to the neutral
gauge boson $i$ are expressed as
$J_i^{\mu}=\sum_f{\bar\psi}_f\gamma^{\mu}(L_i^f P_L+R_i^f P_R)\psi_f$, with 
$P_{L,R}=(1\mp\gamma_5)/2$ the projectors
onto the left- and right-handed fermion helicity states.
With these definitions, the SM couplings are
\begin{equation}
R_{\gamma}^{f}=Q_f;\qquad L_{\gamma}^f=Q_f;\qquad R_Z^{f}=-Q_f s_W^2;
\qquad L_Z^f=I_{3L}^f-Q_f s_W^2,\label{smcoupl}\end{equation}
where $Q_f$ are fermion electric charges, and the couplings in 
Eq.~(\ref{amplit}) are normalized as
\begin{equation}
g_L^f=\frac{g_Z}{e}\,L_Z^f,\qquad g_R^f=\frac{g_Z}{e}\,R_Z^f, \qquad
{g^\prime}^f_L=\frac{g_{Z^\prime}}{e}\,L_{Z^\prime}^f, \qquad
{g^\prime}^f_R=\frac{g_{Z^\prime}}{e}\,R_{Z^\prime}^f.
\label{left}
\end{equation}
In what follows, we will limit ourselves to a few representative models
predicting new gauge heavy bosons. Specifically, models inspired by 
GUT inspired scenarios, superstring-motivated ones, and those 
with Left-Right symmetric origin \cite{Rizzo}. These are the $\chi$ model 
occurring in the breaking $SO(10)\to SU(5)\times U(1)_\chi$, the 
$\psi$ model originating in $E_6\to SO(10)\times U(1)_\psi$, and the $\eta$ 
model which is encountered in superstring-inspired models in which 
$E_6$ breaks directly to a rank-5 group. As an example of Left-Right model, 
we consider the particular value $\kappa=g_R/g_L=1$, corresponding to 
the most commonly considered case of Left-Right Symmetric Model (LR). For 
all such grand-unified $E_6$ and Left-Right models the $Z^\prime$ gauge 
coupling in (\ref{notation}) is $g_{Z^\prime}=g_Zs_W$ \cite{Rizzo}. 
\par 
As they are constrained from present low-energy data \cite{Langacker1} 
and from recent data from the Tevatron \cite{Tevatron}, new vector boson 
effects at the LC 
are expected to be quite small and therefore should be disentangled from 
the radiative corrections to the SM Born predictions for the cross section. 
To this aim, in our numerical analysis we follow the strategy of 
Refs.~\cite{Altarelli2}-\cite{Hollik}, in particular we use the improved 
Born approximation accounting for the electroweak one-loop corrections.

\section{Model independent $Z^\prime$ search and discovery limits}

According to Eqs.~(\ref{s+}), (\ref{s-}) and (\ref{tildesigma}), by the 
measurements of $\sigma_+$ and $\sigma_-$
for the different initial electron beam polarizations one determines the
cross sections related to definite helicity amplitudes $A_{\alpha\beta}$. 
From Eq.~(\ref{amplit}), one can observe that the $Z^\prime$ manifests itself 
in these amplitudes by the combination of the product of couplings 
$g^{\prime e}_\alpha\hskip 2pt g^{\prime f}_\beta$ with the propagator
$\chi_{Z^\prime}$. In the situation $\sqrt s\ll M_{Z^\prime}$ we shall 
consider here, only the interference of the SM term with the $Z^{\prime}$
exchange is important and the deviation of each helicity cross section
from the SM prediction is given by
\begin{equation}
\Delta\sigma_{\alpha\beta}\equiv
\sigma_{\alpha\beta}-\sigma_{\alpha\beta}^{SM}=
N_C\, \sigma_{\rm pt}\, 2 \,
{\rm Re}\, \left[\left(Q_e\, Q_f+g_{\alpha}^e\, g_{\beta}^f\,\chi_Z\right)\cdot
\left({g^{\prime}}^e_{\alpha}\, {g^{\prime}}^f_{\beta}
\, \chi_{Z^{\prime}}^{\ast}\right)\right].
\label{deltasig}\end{equation}
As one can see, $\Delta\sigma_{\alpha\beta}$ depend on the same kind of 
combination of $Z^\prime$ parameters and, correspondingly, each such 
combination can be considered as a single `effective' nonstandard
parameter. Therefore, in an analysis of experimental data for
$\sigma_{\alpha\beta}$ based on a $\chi^2$ procedure, a one-parameter fit is
involved and we may hope to get a slightly improved sensitivity to the 
$Z^\prime$ with respect to other kinds of observables.
\par 
As anticipated, in the case of no observed deviation one can evaluate in a 
model-independent way the sensitivity of process (\ref{proc}) to the 
$Z^\prime$ parameters, given the expected experimental accuracy on $\sigma_+$ 
and $\sigma_-$. It is convenient to introduce the general parameterization of 
the $Z^\prime$-exchange interaction used, e.g., in Refs.~\cite{Leike,Pankov1}:
\begin{equation}
\label{coupl}
G^f_L=L^f_{Z^\prime}\, \sqrt{\frac{g^2_{Z'}}{4\pi}\,
\frac{M^2_Z}{M^2_{Z'}-s}},
\qquad
G^f_R=R^f_{Z^\prime}\, \sqrt{\frac{g^2_{Z'}}{4\pi}\,
\frac{M^2_Z}{M^2_{Z'}-s}}.
\end{equation}
An advantage of introducing the `effective' left- and right-handed couplings 
of Eq.~(\ref{coupl}) is that the bounds can be represented on a 
two-dimensional `scatter plot', with no need to specify particular values of 
$M_{Z^\prime}$ or $s$.
\par 
Our $\chi^2$ procedure defines a $\chi^2$ function for any observable $\cal O$: 
\begin{equation}
\label{Eq:chisq}
\chi^2
=\left(\frac{\Delta{\cal O}}{\delta{\cal O}}\right)^2, 
\label{chi2}
\end{equation}
where $\Delta{\cal O}\equiv{\cal O}(Z^\prime)-{\cal O}(SM)$ and 
$\delta{\cal O}$ is the expected uncertainty on the considered
observable combining both statistical
and systematic uncertainties. The domain allowed to the $Z^\prime$ parameters 
by the non-observation of the deviations $\Delta{\cal O}$ within the accuracy 
$\delta{\cal O}$ will be assessed by imposing $\chi^2<\chi^2_{\rm crit}$, where 
the actual value of
$\chi^2_{\rm crit}$ specifies the desired `confidence' level. The numerical
analysis has been performed by means of the program ZEFIT, adapted to the
present discussion, which has to be used along with
ZFITTER \cite{zfitter}, with input values $m_{top}=175$~GeV and
$m_H=300$~GeV.
\par 
In the real case, the longitudinal polarization of the beams will not exactly 
be $\pm 1$ and, consequently, instead of the pure helicity cross section, the 
experimentally measured $\sigma_\pm$ will determine the linear combinations 
on the right hand side of Eqs.~(\ref{s+}) and (\ref{s-}) with 
$\vert P_e\vert$ (and $\vert P_{\bar e}\vert$) less than unity. Thus, 
ultimately, the separation of $\sigma_{RR}$ from $\sigma_{LL}$ will be 
obtained by solving the linear system of two equations corresponding to the 
data on $\sigma_+$ for, e.g., both signs of the electron longitudinal
polarization. The same is true for the separation of $\sigma_{RL}$ and 
$\sigma_{LR}$ using the data on $\sigma_-$. 
\par 
In the `linear' approximation of Eq.~(\ref{deltasig}), and with 
$M_{Z^\prime}\gg\sqrt s$, the constraints from the condition 
$\chi^2<\chi^2_{\rm crit}$ can be directly expressed in terms of 
the effective couplings (\ref{coupl}) as:
\begin{equation}
\label{bounds}
\vert G_\alpha^{e}G_\beta^{f}\vert<
\frac{\alpha_{e.m.}}{2}\sqrt{\chi^2_{crit}}\,
\left(\frac{\delta\sigma^{SM}
_{\alpha\beta}}{\sigma^{SM}_{\alpha\beta}}\right)
\vert A_{\alpha\beta}^{SM}\vert\frac{M_{Z}^2}{s}.
\end{equation}
We need to evaluate the expected uncertainties $\delta\sigma_{\alpha\beta}$. 
To this aim, starting from the discussion of $\sigma_+$, we consider the 
solutions of the system of four equations corresponding to $P_e=\pm P$ and 
$P_{\bar e}=0$ in Eqs.~(\ref{s+}) and (\ref{s-}):
\begin{eqnarray}
\label{SLL}
\sigma_{\rm LL}
&=&\frac{1+P}{P}\sigma_{+}(-P)-\frac{1-P}{P}\sigma_{+}(P), \\
\label{SRR}
\sigma_{\rm RR}
&=&\frac{1+P}{P}\sigma_{+}(P)-\frac{1-P}{P}\sigma_{+}(-P), \\
\label{SLR}
\sigma_{\rm LR}
&=&\frac{1+P}{P}\sigma_{-}(-P)-\frac{1-P}{P}\sigma_{-}(P), \\
\label{SRL}
\sigma_{\rm RL}
&=&\frac{1+P}{P}\sigma_{-}(P)-\frac{1-P}{P}\sigma_{-}(-P).
\end{eqnarray}

From these relations, adding the uncertainties, e.g. $\delta\sigma_+(\pm P)$ 
on $\sigma_+(\pm P)$ in quadrature, $\delta\sigma_{RR}$ has the form 
\begin{equation}
\label{stat}
\delta\sigma_{RR}=\sqrt{
\left(\frac{1+P}{P}\right)^2\left(\delta\sigma_{+}(P)\right)^2+
\left(\frac{1-P}{P}\right)^2\left(\delta\sigma_{+}(-P)\right)^2},
\end{equation}
and $\delta\sigma_{LL}$ can be expressed quite similarly. Also, we combine 
statistical and systematic uncertainties in quadrature. In this case, if 
$\sigma_+(\pm P)$ are directly measured {\it via} the difference 
(\ref{sigma+}) of the integrated cross sections $\sigma_{1+}(\pm P)$ and 
$\sigma_{2+}(\pm P)$, one can see that $\delta\sigma_+^{stat}$ has the 
simple property: 
${\delta\sigma_+(\pm P)^{stat}=
\left(\sigma^{SM}(\pm P)/\epsilon{\cal L}_{int}\right)^{1/2}}$, 
where ${\cal L}_{int}$ is the time-integrated luminosity, $\epsilon$ is the 
efficiency for detecting the final state under consideration and 
$\sigma^{SM}(\pm P)$ is the polarized total cross section. For the 
systematic uncertainty, we use ${\delta\sigma_{+}(\pm P)^{sys}=\delta^{sys}
\left(\sigma_{1+}^2(\pm P)+\sigma_{2+}^2(\pm P)\right)^{1/2}}$, assuming that 
$\sigma_{1+}(\pm P)$ and $\sigma_{2+}(\pm P)$ have the same systematic error 
$\delta^{sys}$. 
One can easily see that $\delta\sigma_{LL}$ can be obtained by changing 
$\delta\sigma_+(P)\leftrightarrow\delta\sigma_+(-P)$ in (\ref{stat}) and that  
the expression for $\delta\sigma_{RL}$ and $\delta\sigma_{LR}$ also follow 
from this equation by  $\delta\sigma_+\rightarrow\delta\sigma_-$. 
\par
Numerically, to exploit Eq.~(\ref{deltasig}) with $\delta\sigma_{\alpha\beta}$ 
expressed as above, we assume the following values for the expected 
identification efficiencies and systematic uncertainties on the various 
fermionic final states \cite{Damerall}: 
$\epsilon=100\%$ and $\delta^{sys}=0.5\%$ for leptons; $\epsilon=60\%$ and
$\delta^{sys}=1\%$ for $b$ quarks; $\epsilon=35\%$ and $\delta^{sys}=1.5\%$
for $c$ quarks. Also, $\chi^2_{crit}=3.84$ as typical for 95\% C.L. with a
one-parameter fit. We take $\sqrt s=0.5$ TeV and a one-year run with 
${\cal L}_{int}=50\,fb^{-1}$. For 
for polarized beams, we assume 1/2 of the total integrated
luminosity quoted above for each value of the electron polarization,
$P_e=\pm P$. Concerning polarization, in the numerical analysis presented
below we take three different values, $P=$1, 0.8 and 0.5,
in order to test the dependence of the bounds on this variable.
\par 
As already noticed, in the general case where process (\ref{proc}) depends on 
all four independent $Z^\prime ff$ couplings, only the products 
$G_R^eG_R^f$ and $G_L^eG_L^f$ can be constrained by the $\sigma_+$ 
measurement {\it via} Eq.~(\ref{deltasig}), while the products $G_R^eG_L^f$ 
and $G_L^eG_R^f$ can be analogously bounded by $\sigma_-$. The exception is 
lepton pair production ($f=l$) with ($e-l$) universality of $Z^\prime$ 
couplings, in which case $\sigma_+$ can individually constrain either $G_L^e$ 
or $G_R^e$. Also, it is interesting to note that such lepton universality 
implies $\sigma_{RL}=\sigma_{LR}$ and, accordingly, for $P_{\bar e}=0$ 
electron polarization drops from Eq.~(\ref{s-}) which becomes equivalent 
to the unpolarized one, with {\it a priori} no benefit from polarization. 
Nevertheless, the uncertainty in Eq.~(\ref{stat}) still depends on the 
longitudinal polarization $P$. The 95\% C.L. upper bounds on the products of 
lepton couplings (without assuming lepton universality) are reported in the 
first three rows of Table~1. 
\begin{table}
\centering
\caption{95\% C.L. model-independent upper limits at LC 
with {${E_{c.m.}=0.5}$ TeV}. For polarized beams, we take 
${\cal L}_{int}=25\, fb^{-1}$ for each possibility of the electron 
polarization, $P_e=\pm P$.}
\medskip
\begin{tabular}{|c|c|c|c|c|c|}
\hline
\multicolumn{2}{|c|}{}
&&&& \\
\multicolumn{2}{|c|}{couplings}  &
$\vert G^{e}_{R}G^{f}_{R}\vert ^{1/2}$
& $\vert G^{e}_{L}G^{f}_{L}\vert ^{1/2}$ &
$\vert G^{e}_{R}G^{f}_{L}\vert ^{1/2}$ &
$\vert G^{e}_{L}G^{f}_{R}\vert ^{1/2}$ \\
\multicolumn{2}{|c|}{}
  &($10^{-3}$) &($10^{-3}$) &($10^{-3}$) & ($10^{-3}$)\\
\hline
\multicolumn{2}{|c|}{observables}
 &${\sigma_{RR}}$&${\sigma_{LL}}$&
 ${\sigma_{RL}}$&${\sigma_{LR}}$ \\
\hline
process & $P$ & & & & \\
\hline
$e^+e^-\to l^+l^-$ & 1.0 & 2.1 & 2.1 & 3.0 & 3.2
\\ \hline
$e^+e^-\to l^+l^-$ & 0.8 & 2.3 & 2.3 & 3.3 & 3.4
\\ \hline
$e^+e^-\to l^+l^-$ & 0.5 & 2.7 & 2.7 & 3.9 & 4.0
\\ \hline
\hline
$e^+e^-\to{\overline{b}}b$ & 1.0 & 1.9 & 2.0 & 2.5 & 4.6
\\ \hline
$e^+e^-\to{\overline{b}}b$ & 0.8 & 2.2 & 2.1 & 2.8 & 4.8
\\ \hline
$e^+e^-\to{\overline{b}}b$ & 0.5 & 3.0 & 2.3 & 3.7 & 5.7
\\ \hline
\hline
$e^+e^-\to{\overline{c}}c$ & 1.0 & 2.3 & 2.6 & 4.1 & 3.9
\\ \hline
$e^+e^-\to{\overline{c}}c$ & 0.8 & 2.5 & 2.7 & 4.5 & 4.1
\\ \hline
$e^+e^-\to{\overline{c}}c$ & 0.5 & 3.2 & 3.0 & 5.5 & 4.6
\\ \hline
\end{tabular}
\label{tab:tab1}
\end{table}

For quark-pair production ($f=c\,,b$), where in general
${\sigma_{RL}\neq\sigma_{LR}}$ due to the appearance of different 
fermion couplings, the analysis takes into account the reconstruction
efficiencies and the systematic uncertainties previously introduced, and in 
Table~1 we report the 95\% C.L. upper bounds on the relevant products of 
couplings.
\begin{figure}[h]
 \epsfxsize=9cm
\centerline{\epsfbox{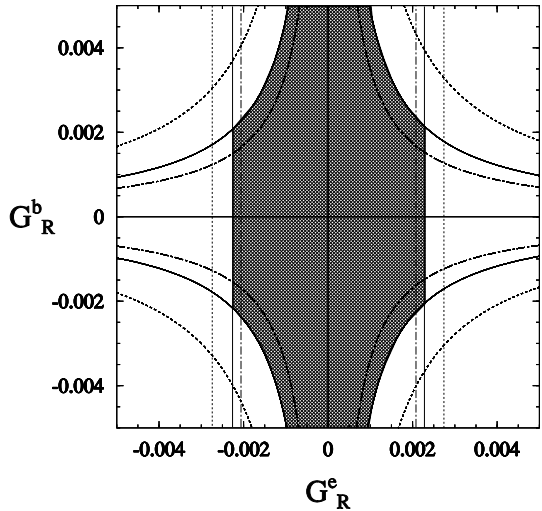}}
\caption{
95\% C.L. upper bounds on the model independent $Z^\prime$ couplings
in the plane {$(G^e_R, G^b_R)$} determined by $\sigma_{RR}$.
The areas enclosed by vertical straight lines are obtained from the process
$e^+e^-\to l^+l^-$, while those enclosed between hyperbolas
are from $e^+e^-\to\bar{b}b$ at ${\cal L}_{int}=50~\mbox{fb}^{-1}$ and
$\sqrt{s}=500$ GeV. The dot-dash, solid and dotted contours 
are obtained at $P=1,\ 0.8,\ 0.5$, respectively.
The shaded region is derived from the combination of
$e^+e^-\to l^+l^-$ and $e^+e^-\to\bar{b}b$ at $P=0.8$.
}
\label{fig1}
\end{figure}
\par 
Also, for illustrative purposes, in Fig.~1 we show the 95\% C.L. bounds in the 
plane $(G_R^e,G_R^b)$, represented by the area limited by the four 
hyperbolas. The shaded region is obtained by combining these limits with 
the ones derived from the pure leptonic process with lepton universality.
Thus, in general we are not able to constrain the individual couplings to a 
finite region. On the other hand, there would be the possibility of using 
Fig.~1 to constrain the quark couplings to the $Z^\prime$ to a finite range 
in the case where some finite effect were observed in the lepton-pair channel. 
The situation with the 
other couplings, and/or the $c$ quark, is similar to the one depicted in 
Fig.~1. 
\par  
Table~1 shows that the integrated observables $\sigma_+$ and
$\sigma_-$ are quite sensitive to the indirect $Z^\prime$ effects, with
upper limits on the relevant products $\vert G_\alpha^e\cdot G_\beta^f\vert$
ranging from $2.2\cdot 10^{-3}$ to $4.8\cdot 10^{-3}$
at the maximal planned value $P=0.8$ of the electron
longitudinal polarization. In most cases, the best sensitivity 
occurs for the $\bar{b}b$ final state, while the worst one is for $\bar{c}c$. 
Decreasing the electron polarization from $P=1$ to $P=0.5$ results in worsening 
the sensitivity by as much as 50\%, depending on the final fermion channel.
\par 
Regarding the role of the assumed uncertainties on the observables under 
consideration, in the cases of $e^+e^-\to l^+l^-$ and
$e^+e^-\to\bar{b}b$ the expected statistics are such that the uncertainty
turns out to be dominated by the statistical one, and the results are almost
insensitive to the value of the systematical uncertainty. Conversely, 
for $e^+e^-\to\bar{c}c$ both statistical and systematic uncertainties are  
important. Moreover, as Eqs.~(\ref{s+}) and (\ref{s-}) show, a
further improvement on the sensitivity to the various $Z^\prime$ couplings
in Table~1 would obtain if both initial $e^-$ and $e^+$ longitudinal 
polarizations were available \cite{Accomando}.

\section{Resolving power and model identification}

If a $Z^\prime$ is indeed discovered, perhaps at a hadron machine, it becomes 
interesting to measure as accurately as possible its couplings and mass at the 
LC, and make tests of the various extended gauge models. To assess the accuracy, 
the same procedure as in the previous section can be applied to the 
determination of $Z^\prime$ parameters by simply  
replacing the SM cross sections in Eqs. (\ref{chi2}) and (\ref{stat}) by  
the ones expected for the `true' values of the parameters (namely, the 
extended model ones), and evaluating the $\chi^2$ variation around them in
terms of the expected uncertainty on the cross section.

\subsection{$Z^\prime$ couplings to leptons}
We now examine bounds on the $Z^\prime$ couplings for $M_{Z^\prime}$ fixed at 
some value. Starting from the leptonic process $e^+e^-\to l^+l^-$,
let us assume that a $Z^\prime$ signal is detected by means of the observables
$\sigma_+$ and $\sigma_-$. Using Eqs.~(\ref{SRR}) and (\ref{SLL}), the 
measurement of $\sigma_+$ for the two values $P_e=\pm P$ 
will allow to extract $\sigma_{RR}$ and $\sigma_{LL}$ which, in 
turn, determine independent and separate values for the right- and left-handed 
$Z^\prime$ couplings $R^e_{Z^\prime}$ and $L^e_{Z^\prime}$ (we assume lepton 
universality). The $\chi^2$ procedure determines the accuracy, or the 
`resolving power' of such determinations given the expected experimental 
uncertainty (statistical plus systematic). 

\begin{table}[t]
\centering
\caption{The values of the $Z^\prime$ leptonic and quark chiral couplings
for typical models with $M_{Z^\prime}=1$ TeV and expected 1-$\sigma$ error 
bars from combined statistical and systematic uncertainties, as determined 
at the LC with $E_{c.m.}=0.5$ TeV and $P=0.8$.}
\medskip
\begin{tabular}{|l|c|c|c|c|} \hline
       & $\chi$  & $\psi$ & $\eta$ & LR \\ \hline
             &     &    &    &     \\
$R^e_{Z'}$   & $ 0.204^{+0.042}_{-0.069}$ & $-0.264_{-0.043}^{+0.052}$
             & $-0.333_{-0.035}^{+0.038}$ & $-0.438_{-0.028}^{+0.029}$ \\
             &     &    &    &     \\ \hline
             &     &    &    &     \\
$L^e_{Z'}$   & $ 0.612^{+0.020}_{-0.020}$ & $ 0.264^{+0.042}_{-0.052}$
             & $-0.166_{-0.061}^{+0.102}$ & $ 0.326^{+0.036}_{-0.039}$ \\
             &     &    &    &     \\ \hline \hline
             &     &    &    &     \\
$R^b_{Z'}$   & $-0.612_{-0.111}^{+0.110}$ & $-0.264_{-0.172}^{+0.111}$
             & $0.166^{+0.096}_{-0.075}$ & $-0.874_{-0.138}^{+0.116}$ \\
             &     &    &    &     \\ \hline
             &     &    &    &     \\
$L^b_{Z'}$   & $-0.204_{-0.042}^{+0.040}$ & $0.264^{+0.158}_{-0.103}$
             & $0.333^{+0.230}_{-0.168}$ & $-0.110_{-0.085}^{+0.080}$ \\
             &     &    &    &     \\ \hline \hline
             &     &    &    &     \\
$R^c_{Z'}$   & $0.204^{+0.092}_{-0.090}$ & $-0.264_{-0.207}^{+0.138}$
             & $-0.333_{-0.145}^{+0.114}$ & $0.656^{+0.122}_{-0.104}$  \\
             &     &    &    &     \\ \hline
             &     &    &    &     \\
$L^c_{Z'}$   & $-0.204_{-0.064}^{+0.059}$ & $0.264^{+0.222}_{-0.149}$
             & $ 0.333^{+0.577}_{-0.326}$ & $-0.110_{-0.134}^{+0.106}$  \\
             &     &    &    &     \\ \hline
\end{tabular}
\label{tab:tab2}
\end{table}

In Table~2 we give the resolution on the $Z^\prime$ leptonic couplings for 
the typical model examples introduced in Section~2, with 
$M_{Z^\prime}=1\, {\rm TeV}$. In this regard, one should recall that the 
two-fold ambiguity intrinsic in process (\ref{proc}) does not allow to 
distinguish the pair of values of 
($g^{\prime e}_{\alpha},g^{\prime f}_{\beta}$)  
from the one ($-g^{\prime e}_{\alpha},-g^{\prime f}_{\beta}$), see 
Eq.~(\ref{deltasig}). Thus, the actual sign of the couplings 
$R_{Z^\prime}^e$ and $L_{Z^\prime}^e$ cannot be determined from the data  
(in Table~2 we have chosen the signs dictated by the relevant models). In 
principle, the sign ambiguity of fermionic couplings might be resolved by 
considering other processes such as, e.g., $e^+e^-\to W^+W^-$. 
\par 
Another interesting question is the potential of the leptonic process 
(\ref{proc}) to identify the $Z^\prime$ model underlying the measured 
signal, through the measurement of the helicity cross sections $\sigma_{RR}$ 
and $\sigma_{LL}$. Such cross sections only depend on the relevant leptonic 
chiral coupling and on $M_{Z^\prime}$, so that such resolving power clearly 
depends on the actual value of the $Z^\prime$ mass. In Figs.~2a and 2b we show 
this dependence for the $E_6$ and the $LR$ models of interest here. In these 
figures, the horizontal lines represent the values of the couplings predicted 
by the various models, and the lines joining the upper and the lower ends of 
the vertical bars represent the expected experimental uncertainty at the 
95\% CL. The intersection of the lower such lines with the $M_{Z^\prime}$ 
axis determines the discovery reach for the corresponding model: larger 
values of $M_{Z^\prime}$ would determine a $Z^\prime$ signal smaller than the 
experimental uncertainty and, consequently, statistically invisible. 
Also, Figs.~2a and 2b show the complementary roles of $\sigma_{LL}$ and 
$\sigma_{RR}$ to set discovery limits: while $\sigma_{LL}$ is mostly 
sensitive to the $Z^\prime_\chi$ and has the smallest sensitivity to the 
$Z^\prime_\eta$, $\sigma_{RR}$  provides the best limit for the $Z^\prime_{LR}$
and the worst one for the $Z^\prime_\chi$. 
\begin{figure}[t]
 \epsfclipon
 \epsfxsize=9cm
 \centerline{
 \epsfxsize=9cm
 \epsfbox{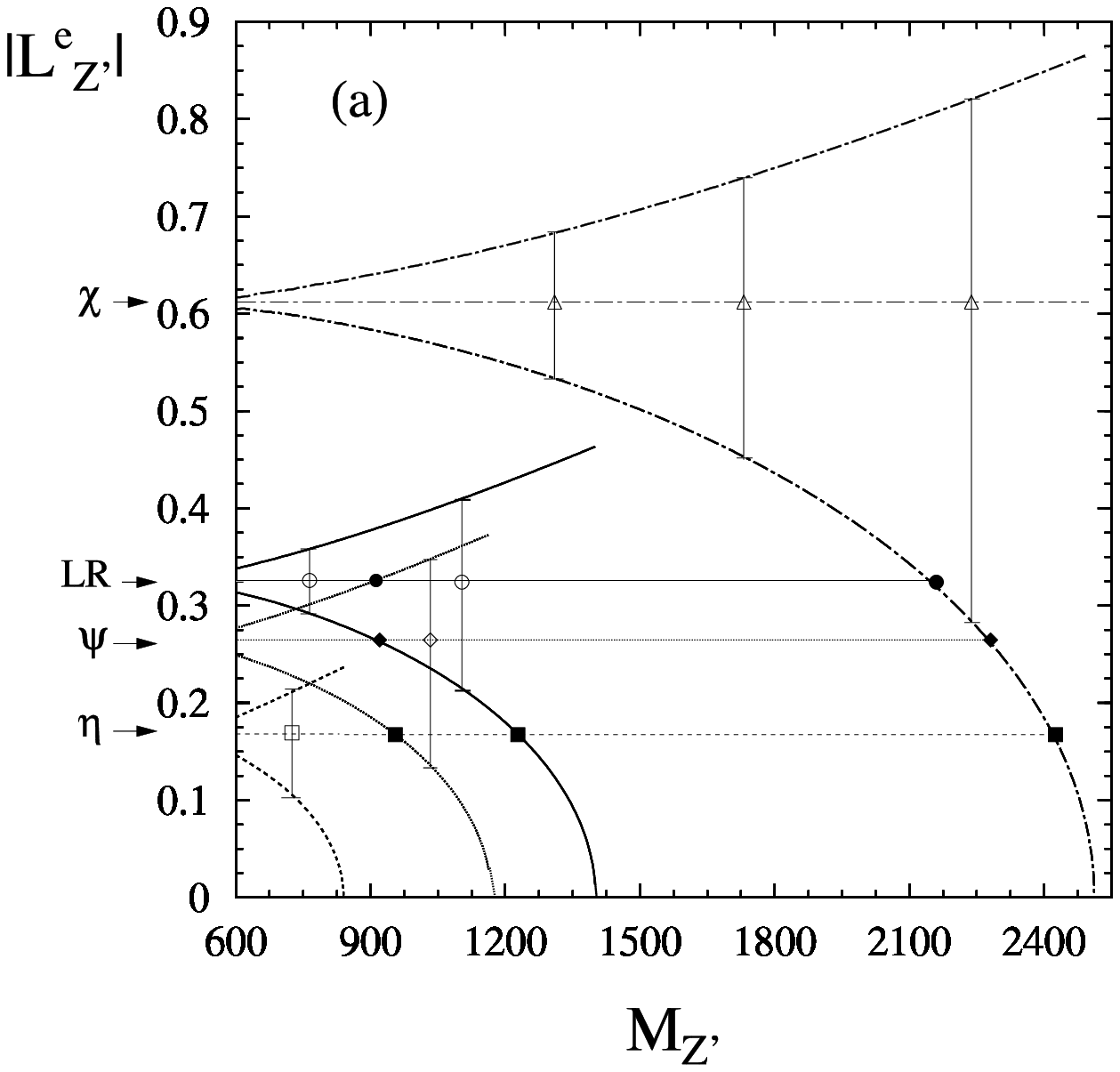}
 \epsfxsize=9cm
 \epsfbox{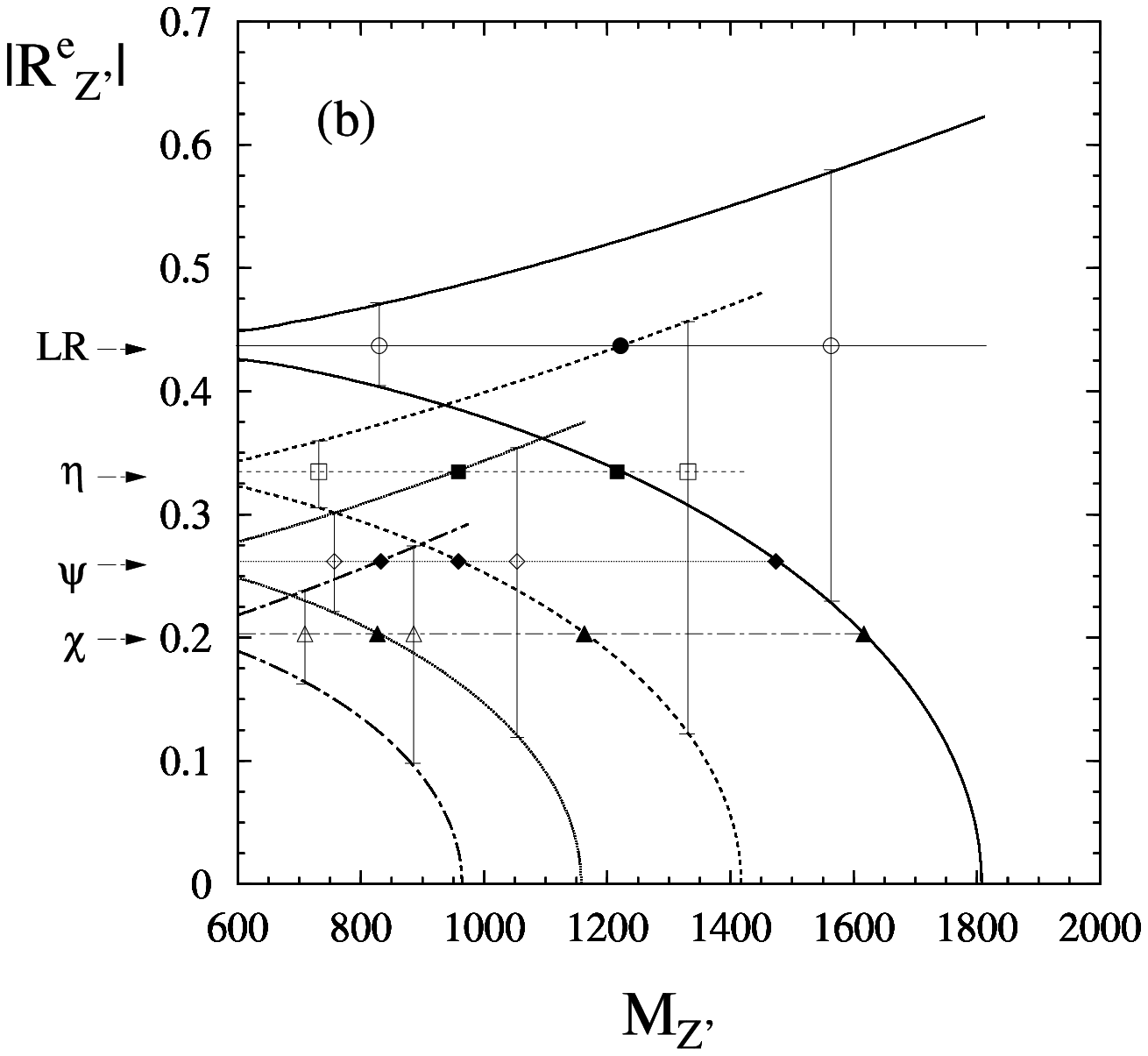}}
 \caption{
Resolution power at 95\% C.L. for the absolute value of the 
leptonic $Z^\prime$ couplings, $\vert L^e_{Z^\prime}\vert$ (a) and 
$\vert R^e_{Z^\prime}\vert$ (b), 
as a function of $M_{Z^\prime}$, obtained 
from $\sigma_{LL}$ and $\sigma_{RR}$, respectively, 
in process $e^+e^-\to l^+l^-$.  The error bars 
combine statistical and systematic uncertainties. 
Horizontal lines correspond 
to the values predicted by typical models. 
}
 \label{fig2}
\end{figure}
 
As Figs.~2a and 2b show, the different models can be distinguished by means of 
$\sigma_\pm$ as long as the uncertainty of the coupling of one model does 
not overlap with the value predicted by the other model. Thus, the
identification power of the leptonic process (\ref{proc}) is determined by the 
minimum $M_{Z^\prime}$ value at which such `confusion region' starts. 
For example, 
Fig.~2a shows that the $\chi$ model cannot be distinguished from the LR, 
$\psi$ and $\eta$ models at $Z^\prime$ masses larger than 2165 GeV, 2270 GeV 
and 2420 GeV, respectively. The identification power for the typical models 
are indicated in Figs.~2a and 2b by the symbols circle, diamond, square and 
triangle. The corresponding $M_{Z^\prime}$ values at 95\% C.L. 
for the typical $E_6$ and LR models are listed in Table~3, where the 
$Z^\prime$ models 
listed in first columns should be distinguished from the ones listed in 
the first row assumed to be the origin of the observed $Z^\prime$ signal. 
For this reason Table~3 is not symmetric.
\begin{table}
\centering
\caption{Identification power of process $e^+e^-\to\bar{f}f$
at 95\% C.L. expressed in terms of $M_{Z^\prime}$ (in GeV) 
for typical $E_6$ and LR models at $E_{c.m.}=0.5$ TeV and
${\cal L}_{int}=25\, fb^{-1}$ for each value of the electron polarization,
$P_e=\pm 0.8$.}\medskip
\begin{tabular}{|c|c|c|c|c||c|c|c|c|} \hline
\multicolumn{1}{|c|}{} &
\multicolumn{4}{|c||}{$\sigma_{RR}$} &
\multicolumn{4}{|c|}{$\sigma_{LL}$} \\ \hline
$e^+e^-\to l^+l^-$ & $\psi$  & $\eta$ & $\chi$ & LR
       & $\psi$  & $\eta$ & $\chi$ & LR \\ \hline
$\psi$ & --- & 960  & 830 & 1470 & --- & 840 & 2270 & 920  \\ \hline
$\eta$ & 950 & ---  & 970 & 1210 & 960 & --- & 2420 & 1220 \\ \hline
$\chi$ & 830 & 1165  & --- &1615 &1170 & 840 & ---  & 1400 \\ \hline
  LR   &1160 & 1220  & 970 & ---  & 915& 840 & 2165 & ---  \\ \hline\hline
$e^+e^-\to\bar{b}b$ & $\psi$  & $\eta$ & $\chi$ & LR
       & $\psi$  & $\eta$ & $\chi$ & LR \\ \hline
$\psi$ & --- & 725  & 1180& 2345 & --- & 710 & 1120 & 940  \\ \hline
$\eta$ & 700 & ---  & 1210& 2410 & 750 & --- & 1250 & 750 \\ \hline
$\chi$ & 1175& 1100 & --- & 2130 &1130 & 1140& ---  & 950 \\ \hline
  LR   &1210 & 1100 &1540 & ---  & 940 & 760 & 1370 & ---  \\ \hline\hline
$e^+e^-\to\bar{c}c$ & $\psi$  & $\eta$ & $\chi$ & LR
       & $\psi$  & $\eta$ & $\chi$ & LR \\ \hline
$\psi$ & --- & 865  & 800 & 1740 & --- & 620 & 935  & 800  \\ \hline
$\eta$ & 880 & ---  & 880 & 1580 & 645 & --- & 1035 & 665 \\ \hline
$\chi$ & 760 & 1050 & --- & 1840 &935  & 940 & ---  & 810 \\ \hline
  LR   &1050 & 1280  & 880 & --- & 780 & 685 & 1135 & ---  \\ \hline
\end{tabular}
\label{tab:tab3}
\end{table}

Analogous considerations hold also for $\sigma_{LR}$ and $\sigma_{RL}$. 
These cross sections give qualitatively similar results for the product 
$L^e_{Z^\prime}R^e_{Z^\prime}$, but with weaker constraints because of
smaller sensitivity.

\subsection{$Z^\prime$ couplings to quarks}

In the case of process (\ref{proc}) with ${\bar q}q$ pair production (with 
$q=c,\, b$), the analysis is complicated by the fact that the relevant 
helicity amplitudes depend on three parameters ($g^{\prime e}_\alpha$,
$g^{\prime q}_\beta$ and $M_{Z^\prime}$) instead of two. Nevertheless, there 
is still some possibility to derive general information on the $Z^\prime$ 
chiral couplings to quarks. Firstly, by the numerical procedure 
introduced above one can determine from the measured cross section the products 
of electrons and 
final state quark couplings of the $Z^\prime$, from which one derives 
allowed regions to such couplings in the independent, two-dimensional, planes 
($L^e_{Z^\prime}$,$L^q_{Z^\prime}$) and ($L^e_{Z^\prime}$,$R^q_{Z^\prime}$).
The former regions are determined through $\sigma_{LL}$, and the latter ones 
through $\sigma_{LR}$. As an illustrative example, in Fig.~3 we depict 
the bounds from the process $e^+e^-\to{\bar b}b$ in the 
($L^e_{Z^\prime}$,$L^b_{Z^\prime}$) and ($L^e_{Z^\prime}$,$R^b_{Z^\prime}$) 
planes for the $Z^\prime$ of the $\chi$ model, with 
$M_{Z^\prime}= 1\,{\rm TeV}$. Taking into account the above mentioned two-fold 
ambiguity, the allowed regions are the ones included within the two sets of 
hyperbolic contours in the upper-left and in the lower-right corners of 
Fig.~3. Then, to get finite regions for the quark couplings, one must 
combine the hyperbolic regions so obtained with the determinations of the 
leptonic $Z^\prime$ couplings from the leptonic process (\ref{proc}), 
represented by the two vertical strips. The corresponding shaded areas 
represent the determinations of $L^b_{Z^\prime}$, while the hatched 
areas are the determinations of $R^b_{Z^\prime}$. Notice that, in general, 
there is the alternative possibility of deriving constraints on quark couplings
also in the case of right-handed electrons, namely, from the determinations of
the pairs of couplings ($R^e_{Z^\prime}$,$L^b_{Z^\prime}$) and
($R^e_{Z^\prime}$,$R^b_{Z^\prime}$). However, as observed with regard to the
previous analysis of the leptonic process, the sensitivity to the 
right-handed electron coupling turns out to be smaller than for 
$L^e_{Z^\prime}$, so that the corresponding constraints are weaker. 

\begin{figure}[t]
 \epsfxsize=9cm
\centerline{\epsfbox{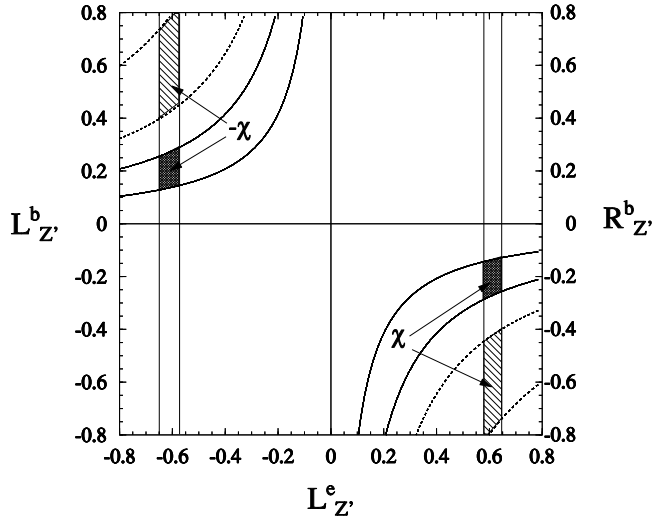}}
\caption{
Allowed bounds at 95\% C.L. on $Z^\prime$ couplings with $M_{Z^\prime}=1$ TeV
($\chi$ model) in the two-dimension planes
($L^e_{Z^\prime}$,$L^b_{Z^\prime}$) and ($L^e_{Z^\prime}$,$R^b_{Z^\prime}$)
obtained from helicity cross sections $\sigma_{LL}$ (solid lines) and
$\sigma_{LR}$ (dashed lines), respectively.
The shaded and hatched regions are derived from the combination of
$e^+e^-\to l^+l^-$ and $e^+e^-\to\bar{b}b$ processes.
Two allowed regions for each helicity cross section correspond to the two-fold
ambiguity discussed in text.
}
\label{fig3}
\end{figure}
\par
The determinations of the $Z^\prime$ couplings with the $c$ and $b$ quarks 
for the typical $E_6$ and LR models with $M_{Z^\prime}=1\,{\rm TeV}$, are given 
in 
Table~2 where the combined statistical and systematic uncertainties are taken 
into account. Furthermore, similar to the analysis presented in Section~4.1 and 
the corresponding Figs.~2a and 2b, we depict in Figs.~4a and 
4b the different models identification power as a function of 
$M_{Z^\prime}$, for the reaction $e^+e^-\to\bar{b}b$ as a representative
example. The model identification power of the $\bar{b}b$ and $\bar{c}c$ pair 
production processes are reported in Table~3.
\begin{figure}[t]
 \epsfclipon
 \epsfxsize=9cm
 \centerline{
 \epsfxsize=9cm
 \epsfbox{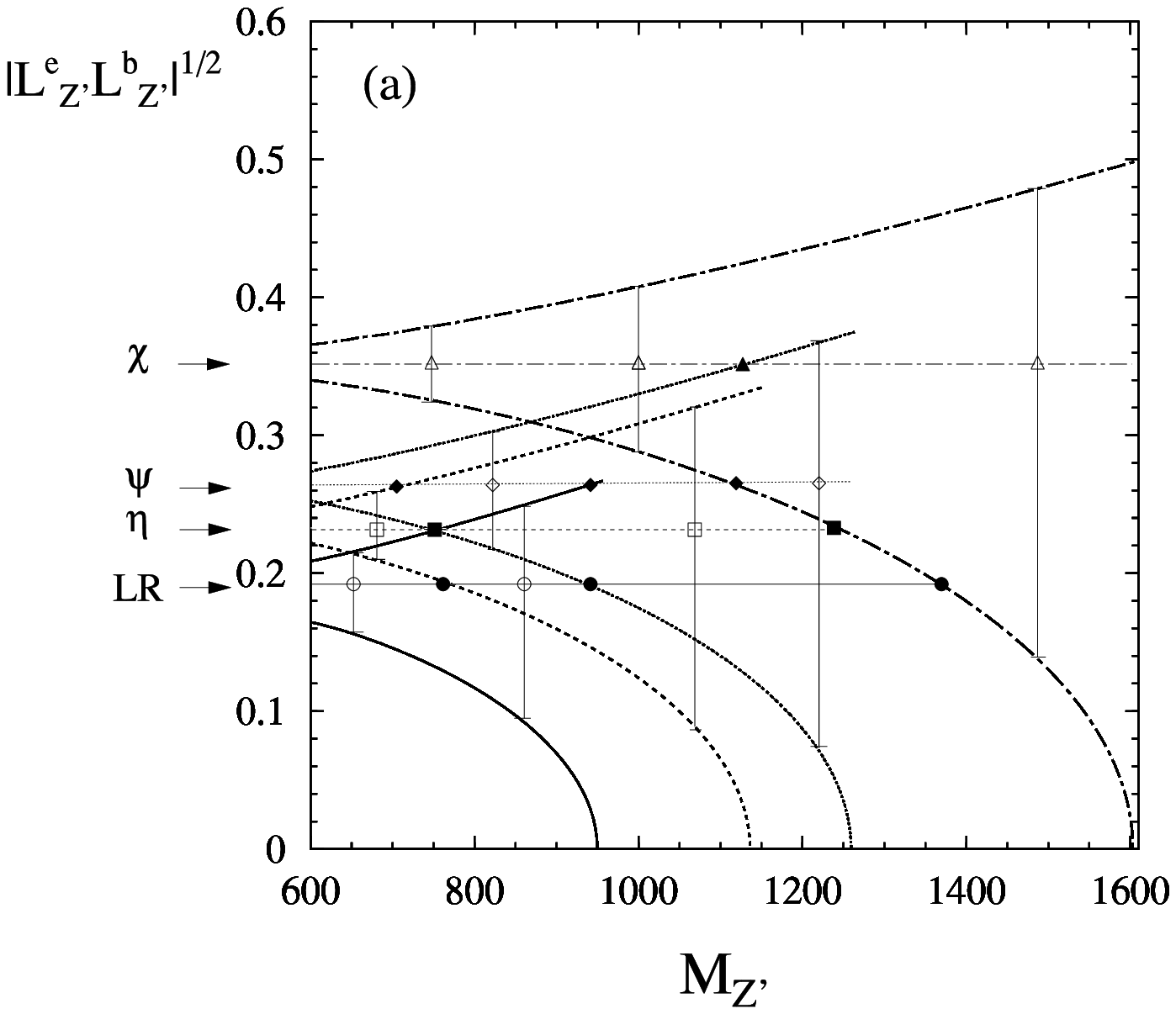}
 \epsfxsize=9cm
 \epsfbox{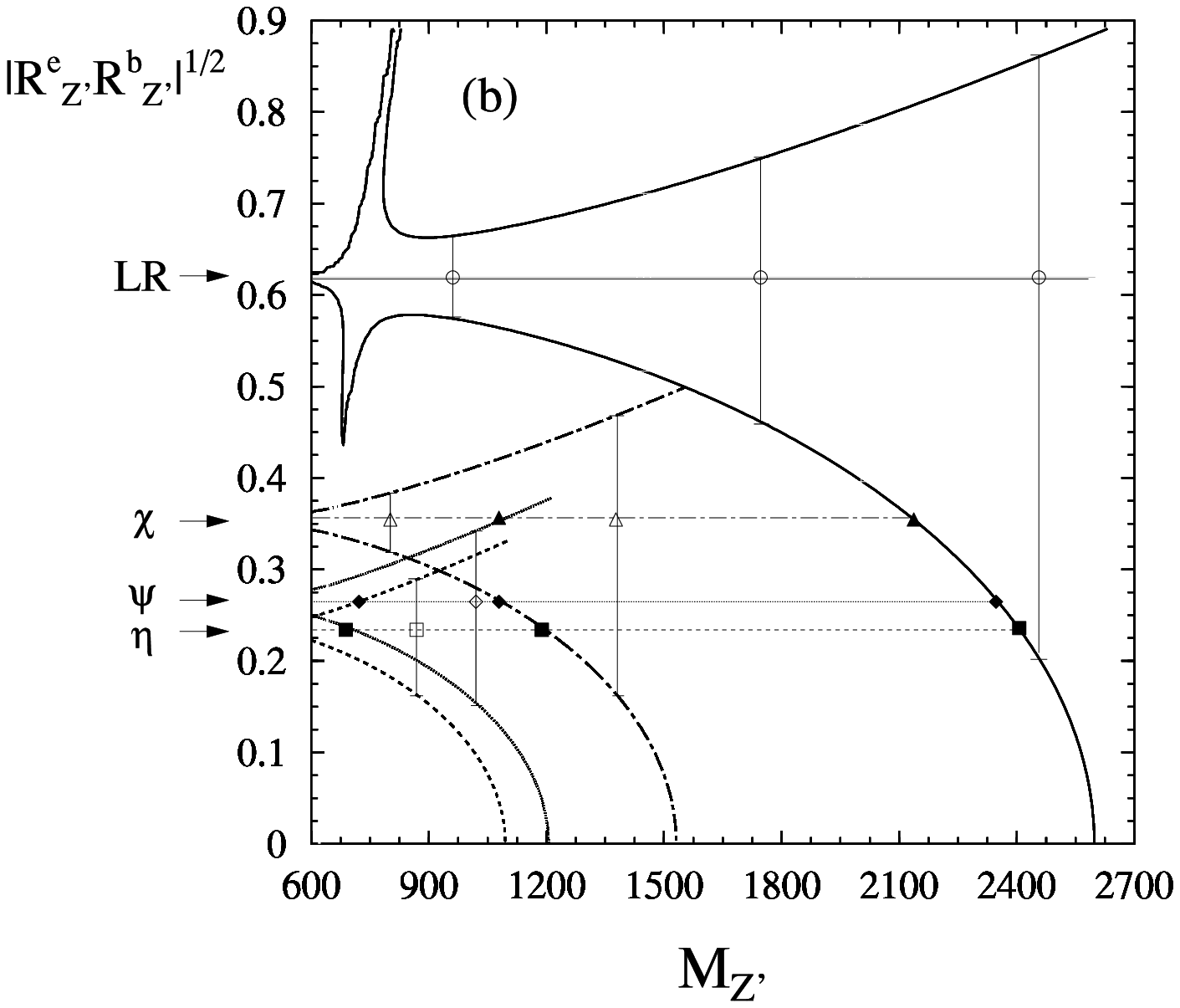}}
 \caption{
Resolution power at 95\% C.L. for 
$\vert{L^e_{Z^\prime}L^b_{Z^\prime}}\vert^{1/2}$ (a)
and $\vert{R^e_{Z^\prime}R^b_{Z^\prime}}\vert^{1/2}$ (b)
as a function of $M_{Z^\prime}$ obtained from
$\sigma_{LL}$ and $\sigma_{RR}$, respectively, in process $e^+e^-\to\bar{b}b$.  The error bars 
combine statistical and systematic errors.  
Horizontal lines correspond to the 
values predicted by typical models.
  }
 \label{fig4}
\end{figure}

\section{Conclusion}

We briefly summarize our findings concerning the $Z^\prime$ discovery limits 
and the models identification power of process (\ref{proc}) {\it via} the 
separate measurement of the helicity cross sections 
$\sigma_{\alpha\beta}$ at the LC, with $\sqrt s=0.5\, {\rm TeV}$ and 
${\cal L}_{int}=25\hskip 2pt fb^{-1}$ for each value $P_e=\pm P$ the electron 
longitudinal polarization. Given the present experimental lower limits on 
$M_{Z^\prime}$, only indirect effects of the $Z^\prime$ can be studied at 
the LC. In general, the helicity cross sections allow to extract  
separate, and model-indpendent, information on the individual `effective' 
$Z^\prime$ couplings ($G_\alpha^e \cdot G_\beta^f$). As depending on the 
minimal number of free parameters, they may be expected to show some 
convenience with respect to other observables in an analysis of the 
experimental data based on a $\chi^2$ procedure. 
\par    
In the case of no observed signal, i.e., no deviation of 
$\sigma_{\alpha\beta}$ from the SM prediction within the experimental accuracy, 
one can directly obtain model-independent bounds on the leptonic chiral 
couplings of the $Z^\prime$ from $e^+e^-\to l^+l^-$ and on the products of 
couplings $G_\alpha^e\cdot G_\beta^q$ from $e^+e^-\to {\bar q}q$ (with 
$l=\mu,\tau$ and $q=c,b$). From the numerical point of view, 
$\sigma_{\alpha\beta}$ are found to just have a complementary role with respect 
to other observables like $\sigma$ and $A_{\rm FB}$.
\par  
In the case $Z^\prime$ manifestations are observed as deviations 
from the SM, with $M_{Z^\prime}$ of the order of {1 TeV}, the role of 
$\sigma_{\alpha\beta}$ is more interesting, specially as regards the problem of 
identifying the various models as potential sources of such non-standard 
effects. Indeed, in principle, they provide a unique  possibility to 
disentangle and extract numerical values for the chiral couplings of 
the $Z^\prime$ in a general way (modulo the aforementioned 
sign ambiguity), avoiding the danger of cancellations, so that 
$Z^\prime$ model predictions can be tested. Data analyses with 
other observables may involve combinations of different coupling constants and  
need some assumption to reduce the number of independent parameters in the 
$\chi^2$ procedure. In 
particular, by the analysis combining $\sigma_{\alpha\beta}(l^+l^-)$ and 
$\sigma_{\alpha\beta}({\bar q}q)$ one can obtain information of the 
$Z^\prime$ couplings with quarks without making assumptions on the values of 
the leptonic couplings. Numerically, as displayed in the previous Sections,for the class of $E_6$ and Left-Right models considered here the couplings 
would be determined to about $3-60\%$ for $M_{Z^\prime}=1\,{\rm TeV}$. 
Of course, the considerations above hold only in the case where the $Z^\prime$ 
signal is seen in all observables. 
Finally, one can notice that for $\sqrt s\ll M_{Z^\prime}$ the 
energy-dependence of the deviations $\Delta\sigma_{\alpha\beta}$ is 
determined by the SM and that, in particular, the definite sign 
$\Delta\sigma_{\alpha\alpha}(l^+l^-)<0$ ($\alpha=L,R$) is typical of the 
$Z^\prime$. 
This property might be helpful in order to identify the $Z^\prime$ as the 
source of observed deviations from the SM in process (\ref{proc}).      

\section*{Acknowledgements}
It is a pleasure to thank N. Paver for the fruitful 
and enjoyable collaboration on the topics covered here.


\end{document}